# Coherent Control of the Waveforms of Recoilless Gamma-Photons

Farit Vagizov[1,2], Vladimir Antonov[3,1], Y.V. Radeonychev[3,1], R. N. Shakhmuratov[2]

and Olga Kocharovskaya[1]

[1] *Department of Physics and Astronomy and Institute for Quantum Studies and Engineering,*

*Texas A&M University, College Station, TX: 7784-4242, USA*

[2] *Kazan Federal University and Kazan Physical Technical Institute of the Russian Academy of Sciences,*

*Kazan 420008, Russia*

[3] *Institute of Applied Physics of the Russian Academy of Sciences, Nizhny Novgorod 603950, Russia*

The concepts and ideas of coherent, nonlinear and quantum optics deeply penetrate into the range of 10-100 kiloelectronvolt (keV) photon energies, corresponding to soft gamma-ray (hard x-ray[*]) radiation. The recent experimental achievements in this frequency range include demonstration of the parametric down-conversion in the Langevin regime [1], cavity electromagnetically induced transparency [2], collective Lamb shift [3], and single-photon revival in the nuclear absorbing sandwiches [4]. Realization of a single photon coherent storage [5] and stimulated Raman adiabatic passage [6] were recently proposed. Still the number of tools for coherent manipulation of gamma-photon – nuclear ensemble interactions remains rather limited. In this work an efficient method to coherently control the waveforms of gamma-photons has been suggested and verified. In particular, the temporal compression of an individual gamma-photon into coherent ultrashort pulse train has been demonstrated. The method is based on the resonant interaction of gamma-photons with an ensemble of nuclei with modulated frequency of the resonant transition. The frequency modulation, achieved by uniform vibration of the resonant absorber due to the Doppler Effect, results in the time-dependence of the resonant absorption and dispersion, which allow shaping of the incident gamma-photons. The developed technique is expected to give a strong impetus on emerging fields of coherent and quantum gamma-optics, providing a basis for realization of the gamma-photon – nuclear ensemble interfaces and quantum interference effects at the nuclear gamma-ray transitions.

---

[*] It is a historic tradition to call a radiation in this range x-ray radiation when it is produced by electron motion and to call it gamma-ray radiation if it is produced by nuclear transitions.

Quantum optics is the field of research dealing with interactions of quanta of electromagnetic radiation with quantum transitions of matter. It provides the basis for new fast growing fields of quantum cryptography, communication, and information. So far the experiments in these fields have been implemented either with microwave or optical photons, interacting with atomic electron transitions, and typically required cryogenic temperatures. The gamma-photons in the range of 10-100keV and the corresponding nuclear quantum transitions are the most suitable for realization of such experiments due to nearly 100% detector efficiency, extremely high Q-factor (~$10^{12}$ for 14.4keV transition in $^{57}$Fe) of recoilless nuclear transitions even at room temperature, existence of radioactive materials (representing themselves the natural sources of single gamma-photons) and the cascade scheme of radiative decay of some radioactive sources (Fig.1a), allowing one to study the photon temporal shape via time-delayed coincidence measurement technique [7]. Moreover, the gamma-photons have important potential advantages over the microwave and optical photons for applications in cryptography, communication and information due to extremely high (sub-Å) spatial resolution, high transparency of many materials in this frequency range, and potentially high capacity of the information channels. In particular, the fundamental limitation on the way to miniaturization of photonic circuits settled by the diffraction limit and determined by the photon wavelength (~1μm in optics) would not be an issue for gamma-photons. However, the methods to coherently control the temporal waveforms of single photons (well developed in optics due to the presence on high finesse cavities and bright coherent sources of radiation) are rather limited in gamma-optics (although the emerging state-of-the-art facilities, such as hard X-ray FELs [8,9] and recently built nearly 100% reflecting mirrors [10] will provide an impetus for the development of such techniques).

We propose an efficient method to coherently control the temporal shape of gamma-photons in optically thick uniformly vibrated resonant recoilless absorber verified by the proof-of-principle experiments with radioactive $^{57}$Co:Rh source of single gamma-photons and a stainless steel absorber with natural abundance of $^{57}$Fe isotope (Fig.1).

The basic idea is the following. The vibration of an absorber leads to modulation of the resonant $|1\rangle$-$|2\rangle$-transition frequency (Fig.1(b)) with respect to the frequency of the incident photons and vice versa due to the Doppler Effect. In the case of harmonic vibration with amplitude *R*, providing the modulation index $p=2\pi R/\lambda>1$ ($\lambda$ is the photon wavelength), and vibration frequency $\Omega/2\pi$ exceeding $|1\rangle$-$|2\rangle$-transition linewidth, the spectrum of incoming photons is enriched by the well-separated (with frequency $\Omega$) spectral sidebands. We emphasize the crucial role of phases of the spectral components in formation of the temporal waveforms (unlike the Mössbauer absorption spectra measurements, where information about phases is typically lost). In particular, a constant phase difference between the equidistant neighboring spectral components corresponds to a sequence of pronounced bandwidth-limited pulses and determines the phase of this sequence at the photon front edge. The pulse repetition rate is equal to the modulation frequency, and the pulse duration is defined by the total number of spectral components controlled by vibration amplitude.

It is worth noting that the frequency modulation of the recoilless emission via vibration (typically, of the radioactive source) has being intensively used since late 1970s to study coherent effects such as dynamical beats, quantum beats, and Mössbauer transients ([11] and references therein). These effects were proved to be useful spectroscopic tools for determination of hyperfine interaction parameters (such as isomer shifts, recoilless fractions, quadrupole splitting). They were used also to calibrate mechanical displacement of a source relative to absorber and to determine their linewidths and recoilless fractions. More recently, it was suggested to produce ultrashort gamma-pulses using a vibrated source, emitting frequency modulated radiation, and a far-off-resonant absorber, compensating the group-velocity dispersion [12]. Correct in principle, that proposal is difficult to implement experimentally, because weak nonresonant interaction requires rather long sample, leading to large off-resonant losses. Use of the resonant absorber resolves this problem. Besides, it is easier to provide uniform oscillations of all nuclei in a thin film of absorber than in the radiative source.

The discussed method to produce ultrashort gamma-ray pulses essentially represents itself gamma-ray–nuclei implementation of the general approach based on variation of parameters of the resonant quantum transition [13–15]. While in the case of electronic transitions in atoms modulation of the resonant transition frequency is due to quasi-static Stark effect caused by a strong laser field, in the case under consideration it is due to the Doppler Effect caused by mechanical vibration. This resonant parametric technique, unlike the nonlinear optics techniques, such as high harmonic generation [16] or stimulated Raman scattering [17], does not require high intensity of incoming radiation and hence can be implemented even in a single-photon regime.

Observation of the single-photon wavefom shaped by the vibrated absorber is based on the time-delayed coincidence counts of sequentially emitted 122keV and 14.4keV photons. Registration of 122keV photon defines the initial moment $t_0$ of 14.4keV photon emission due to radiative decay of the excited state $|b\rangle$ of the radioactive source (Eq.(2) in Supplement), and triggers the measurement of detection delay $\tau = t - t_0$ of 14.4keV photon at the exit of the absorber. Multiple measurements of $\tau$ give the coincidence count rate of 14.4keV photon, $N(\tau)$, corresponding to a single-photon waveform, $|\Psi(\tau)|^2$ [4,7,18–20]. Transmission through the vibrated absorber results in reshaping of the exponential waveform of the incident photon depending on the vibration phase $\vartheta_0$ at time $t_0$, $\vartheta_0 = \Omega t_0$. However, the stochastic nature of $t_0$ leads to random phase $\vartheta_0$, while in order to reveal the effect of coherent vibration on the single-photon waveform, one has to accumulate the photons in the same phase $\vartheta_0$. This problem is resolved by collecting the coincidence counts only within the short intervals $dt$ in vicinity of the moments $t_0^{(n)} = (\vartheta_0 + 2\pi n)/\Omega$ (Fig.2), corresponding to the same vibration phase, $\vartheta_0$. In other words, from a full flow of 14.4keV photons only those are counted, which follow 122keV photons, registered at fixed phase of vibration, $\vartheta_0$ (within an instrumental accuracy $d\vartheta = \Omega dt$). A simplified scheme of the experimental setup is given in Fig.2.

This technique allowed us to observe individual gamma-photon temporally compressed into decaying train of ultrashort pulses, an order of magnitude shorter than the total duration of the photon (Fig.3(b)), as well as into the double-humped pulse (Fig.3(c)).

The clear physical picture of such drastic waveform transformation is revealed by the spectral analysis. The quasi-monochromatic spectrum (Fig.3(a)-inset) of the incoming photon (Fig.3(b), red dashed curve) in the laboratory reference frame is "seen" by the vibrating nuclei of the absorber as a comb of equidistant spectral components separated by the vibration frequency $\Omega$ (Fig.3(a)) with amplitudes and phases determined by the Bessel functions (Eqs.(4),(5) in Supplement). Under the chosen parameter values (Fig.3-caption), there are seven major spectral components. Elimination of the "+1" sideband leads to phase matching between five of the remaining components with phase difference of $\pi n$, $n \neq 0$ (three brown squares, red circle, and green triangle in Fig.3(a)), resulting in the decaying train of pulses in Fig.3(b). On the other hand, elimination of the "-1" sideband would lead to phase matching between five of the remaining components with phase difference of $2\pi n$ (three green triangles, red circle, and brown square in Fig.3(a)), resulting in the decaying train of pulses in Fig.3(b)-inset, which differs from the former case by the phase of the waveform at the front part of the photon. Elimination of either spectral component can be achieved by tuning it into the resonance with the absorbing transition via the proper choice of constant velocity of an emitter with respect to absorber. The efficiency of transformation of the incident photon into the pulse train is rather high since the energy of only one sideband is lost. Due to constructive interference of sidebands, the peak intensities of pulses exceed the incident photon intensity. Even higher efficiency and peak pulse intensities can be achieved if, instead of resonant suppression of either phase mismatched sideband, its phase is shifted by $\pi$ due to proper detuning between the emitter and absorber spectral lines. Such phase shift can be realized without undesirable nonresonant absorption in a highly enriched by $^{57}$Fe stainless steel film.

The number and shape of pulses, pulse duration and repetition rate within the single-photon waveform can be controlled by variation of frequency and amplitude of vibration, vibration phase $\vartheta_0$, optical depth of the absorber, as well as detuning between the emitter and absorber spectral lines, allowing one to produce gamma photons with various waveforms. A particular example of the experimentally observed single gamma-photon waveform, possessing by two sharp peaks of nearly equal amplitudes, is shown in Fig.3(c).

Removing of the "D1" 122keV-photon detector from the experimental setup in Fig.2 and using the "Gate in" signal from the generator "G" as a "Start" signal for "TAC" correspond to averaging of 14.4keV-photon count rate over the moment $t_0$ (Eq.(13) in Supplement). As a result, the full flow of shaped 14.4keV photons is summing up into a train of nanosecond pulses (Fig.4). An increase of the vibration frequency at fixed vibration amplitude results in proportional shortening of both pulse duration and repetition period due to widening of the output spectrum via increase of separation between the spectral components (Fig.4(a) vs. Fig.4(b) and Fig.4(c) vs. Fig.4(d)). An increase of the vibration amplitude at fixed vibration frequency results in pulse shortening without change of the repetition period due to increasing number of sidebands at the cost of pulse shape distortion because of appearance of the phase-mismatched components (Fig.4(a) vs. Fig.4(c) and Fig.4(b) vs. Fig.4(d)).

This set up represents itself a table-top source of the ultra-short gamma-pulse trains with the unique coherent properties. Namely, their spectrum consists of extremely narrow (~1MHz) equidistant phase-matched components and the pulses are nearly bandwidth-limited. Such trains of pulses are well suited for realization of quantum interference effects at the nuclear transitions (such as electromagnetically induced transparency [21, 22] and modulation induced transparency [23]). They are promising for the broadband high-spectral-resolution spectroscopy of nuclei in solids in the 10-100keV frequency range and can be used also to study the dynamical processes in complex solid compounds by means of time-resolved gamma-ray spectroscopy and diffraction with resolution up to 100ps. The characteristics of the gamma-pulse sequences, including the shape of the pulses, can be widely controlled both in the multiphoton and single photons regimes. Coherent control of the single-photon waveforms opens the fascinating possibilities for applications of gamma-photons and gamma-photon–nuclear-ensemble interfaces (in particular, quantum memories [24]) in quantum communication and information processing [24,25].

**Supplementary Information** is linked to the online version of the paper at www.nature.com/nature.

**Acknowledgements** We acknowledge the support from NSF (grant No. 0855688), RFBR (grants No. 13-02-00831, No. 12-02-12101, No. 12-02-31325, No. 12-02-33074, and No. 12-02-00263), and Ministry of Science and Education of Russian Federation (contracts No. 11.G34.31.0011, No. 07.514.11.4162, and No. 8520). V. Antonov acknowledges support from the ''Dynasty'' Foundation.

**Author Contributions**: F.V. performed the experiments. V. A., Y.R. and R. S. developed the theoretical description. V.A. determined the optimal parameters for the experiments and provided the theoretical fit to experimental data. Y.R. suggested the technique for observation of the single-photon waveforms. R.S. obtained analytical solutions for some limiting cases. O.K. suggested the idea, coordinated the efforts and wrote the paper. All authors discussed the results and edited the manuscript.

**Author Information**: Reprints and permissions information are available at www.nature.com/reprints. The authors declare no competing financial interests. Correspondence and requests for materials should be addressed to kochar@physics.tamu.edu

**Figure legends**

**Fig.1.** (a): The energy scheme of the $^{57}$Co radioactive decay. The $^{57}$Co nuclide in the energy state $|d\rangle$ decays with half-life decay time $T_{1/2}$=272 days via electron capture, producing $^{57}$Fe nucleus in the excited state $|c\rangle$, which decays with radiative lifetime $T_{r(c)}$=12ns to the first excited state $|b\rangle$ emitting 122keV gamma-photon. The state $|b\rangle$ radiatively decays with lifetime $T_r$=141ns to the ground state $|a\rangle$ emitting 14.4keV photon. (b): The 14.4keV transition between the ground state $|1\rangle$ and the first excited state $|2\rangle$ of absorbing $^{57}$Fe nucleus with linewidth 1.13MHz.

**Fig. 2**: The scheme of experimental setup for gamma-photon waveform control. The emitter "E" is a foil of radioactive $^{57}$Co:Rh with low activity (~100kBq) to avoid photon overlapping. It is mounted on the holder of the Mössbauer transducer, providing tunable constant velocity, $V_{em}$, of the source relative to the absorber. The absorber "A" is a 25μm thickness stainless steel foil (304, Alfa Aesar) with natural abundance, ~2%, of $^{57}$Fe (corresponding to optical thickness $T_M$=5.18), glued on a 28mm film of polyvinylidene fluoride piezo-transducer (Measurement Specialties, Inc.), which transforms the sinusoidal signal from the radiofrequency generator "G" into the uniform vibration of the absorber. The Time-Amplitude Converter "TAC", operating in the coincidence mode, receives the "Start" pulse, produced by the detector "D1" upon registration of 122keV photon, and the "Stop" pulse, produced by the detector "D2" upon registration of 14.4keV photon, passed through the vibrated absorber. The output voltage from TAC, proportional to time difference between the "Start" and "Stop" pulses, is converted into the time spectrum by the Amplitude-Digital Converter "ADC". At the moments $t_0^{(n)}$, matching the chosen phase of vibration, $\Omega t_0^{(n)} = \vartheta_0 + 2\pi n$, the generator produces also "Gate In" signals, allowing to measure the delayed coincidence counts between the 122keV and 14.4keV photons, emitted only during the short acquisition interval $dt$. In other words, only if 122keV photon is received within the interval between $t_0^{(n)}$ and $t_0^{(n)}+dt$, TAC and ADC start to measure the delayed time of arrival of 14.4keV

photon. As a result of multiple measurements, the output of ADC reproduces a waveform of 14.4keV single gamma photon, passed through and reshaped by the vibrated absorber (Fig.3).

**Fig.3.** (a): Spectrum of the incident 14.4keV-photon waveform in the laboratory reference frame (inset) and in the reference frame of the vibrating absorber, calculated according to Eq.(5) in Supplement (blue solid and red dashed lines are spectral amplitudes and phases, respectively). Central phases of lower-frequency, carrier-frequency, and upper-frequency sidebands are labeled by green triangles, red circle, and brown squares, respectively. The phases of "-2" and "+2" sidebands can be shifted to $2\pi$. The absorber vibration frequency is $\Omega/(2\pi)$=10.2MHz, vibration phase is $\vartheta_0 = 0$, and vibration amplitude is $R$=0.25Å ($p$=1.8). (b): The coincidence count rate of 14.4keV photons normalized to the coincidence count rate at time $t_0$, $N(\tau)/N(t_0)$, passed through the vibrating absorber, as a function of delay time $\tau$, corresponding to the waveform of a single gamma-photon. The "+1" sideband is tuned to the absorber spectral line by moving the emitter towards the absorber with constant velocity $V_{em}$=0.88mm/s. The experimental data are blue dots centered at the confidence intervals. Red solid curve is calculated numerically according to Eq.(12) in Supplement. Red dashed curve represents the waveform of the incident photon calculated according to Eq.(3) in Supplement with spectrum shown in (a)-inset. The upper inset shows sinusoidal vibration of the absorber and the acquisition interval, $\Omega dt=\pi/2$, (gray bar) limited by the equipment used. The lower inset shows the single-photon waveform in the case, where the "-1" sideband is tuned to the absorber resonance. (c): The same as in (b) under the following parameters: absorber vibration frequency is $\Omega/(2\pi)$=2.6MHz, vibration phase is $\vartheta_0 = 0$, vibration amplitude is $R$=0.9Å ($p$=6.58), emitter is moved towards the absorber with constant velocity $V_{em}$=1.13mm/s, and the acquisition interval is $\pi/8$ (gray bar in the upper inset). The lower inset shows the single-photon waveform under the same parameters, except $V_{em}$=−1.13mm/s.

**Fig.4.** Count rate of 14.4keV photons versus time in the case, where detector "D1" (Fig.2) is removed, and the "Gate In" signal from generator "G", matched with zero phase of vibration, is used for triggering of TAC and ADC for recording time of arrival of 14.4keV photon. The "Stop" signal from detector "D2" upon registration of 14.4keV photon stop their operation. As a result of multiple measurements, intensity of 14.4keV radiation at the output of the vibrated absorber is observed. Constant velocity, $V_{em}$, of the emitter towards the absorber tunes the "+1" sideband to the absorber spectral line (Fig.3(a)). (a): $\Omega/(2\pi)$=5.16MHz, $V_{em}$=0.44mm/s, $R$=0.25Å, ($p$=1.8); (b): $\Omega/(2\pi)$=10.2MHz, $V_{em}$=0.87mm/s, $R$=0.25Å ($p$=1.8); (c): $\Omega/(2\pi)$=5.16MHz, $V_{em}$=0.44mm/s, $R$=0.27Å, ($p$=2); (d): $\Omega/(2\pi)$=10.2MHz, $V_{em}$=0.87mm/s, $R$=0.27Å ($p$=2).

Supplementary information

Transformation of the recoilless gamma-photon in a uniformly vibrated Mössbauer absorber can be considered as follows. The resonant absorber vibrates along the direction of propagation of 14.4keV gamma-photons with amplitude $R$ and frequency $\Omega/(2\pi)$ (Fig.2). The absorber thickness, $L$, meets the condition $L \ll 2\pi V_s/\Omega$ (where $V_s$ is the speed of sound inside the absorber), therefore the coordinate $z$ in the laboratory reference frame is related to the coordinate $z'$ in the reference frame of the absorber as

$$z = z' + R\sin(\Omega t), \quad (1)$$

The electric field of an individual 14.4keV gamma-photon emitted by the radioactive source, $E_r(z,t)$, in the laboratory reference frame can be described as a classical quasi-monochromatic wave [2-4,11,12,18],

$$E_r(z,\tau) = E_0 \theta(\tau - z/c) e^{-(i\omega_r + \Gamma_r)(\tau - z/c) + i\varphi_0}, \quad (2)$$

where $\theta(x)$ is the Heaviside step function, $\tau = t - t_0$ is the time delay in coincidence measurements, and $t_0$ labels the moment a nucleolus of the radioactive source is registered in the excited state $|b\rangle$ (Fig.1(a)), $\Gamma_r = 1/(2T_r)$ ($T_r$ is the radiative lifetime of the state $|b\rangle$), $\omega_r = 2\pi c/\lambda$ is the carrier frequency of the emitted photon, $\lambda$ is the photon wavelength, $c$ is the speed of light, and $\varphi_0$ is the random phase. The propagation distance of the photon form the emitter to the detector meets the relation $z \ll c\tau$. Therefore count rate of the emitted photons before the absorber, calculated from (2), has the form (red dashed lines in Fig.3):

$$N_r(\tau) \sim |E_r(\tau)|^2 = E_0^2 \theta(\tau) e^{-2\Gamma_r \tau}, \quad (3)$$

In the reference frame of the moving absorber the electric field (2) takes the form

$$E_r(z',\tau) = E_0 \theta(\tau - z'/c) e^{-(i\omega_r + \Gamma_r)(\tau - z'/c) + i\varphi_0} \left( J_0(p) + \sum_{n=1}^{\infty} J_n(p) e^{in(\Omega\tau + \vartheta_0)} + J_n(p) e^{-in(\Omega\tau + \vartheta_0 + \pi)} \right), \quad (4)$$

where $\vartheta_0 = \Omega t_0$, $J_n(p)$ is the n-th Bessel function of the first kind, and $p = \frac{2\pi R}{\lambda}$ is the modulation index. Equation (4) results from substitution of (1) into (2) in nonrelativistic approximation, in assumption $\Gamma_r \ll \omega_r$, and in accordance with the transform $e^{ip\sin\alpha} = \sum_{n=-\infty}^{\infty} J_n(p) e^{in\alpha}$ taking into account the relation $J_{-n} = e^{-i\pi n} J_n$. Thus, the electric field, as it is "seen" by moving nuclei, is frequency modulated, or equivalently is a superposition of exponentially decaying quasi-monochromatic components with carrier frequencies $\omega_r \pm n\Omega$ separated by the vibration frequency, $\Omega$. The number and amplitudes of the major sidebands are determined by the amplitude of vibration via $J_n(p)$. The sidebands with amplitudes proportional to the Bessel function magnitudes, $|J_n|$, are symmetrically located in respect to the photon carrier frequency, $\omega_r$. If the modulation index $p<2.4$, then all $J_n$ are positive and the sidebands with lower frequencies (the middle term in the sum) are phase matched with phase difference $\vartheta_0$ determined by the moment $t_0$, while the sidebands with upper frequencies (the last term in the sum) are phase matched with phase difference $\vartheta_0 + \pi$ (see Fig.3(a)).

Fourier transform of Eq.(4) with taking into account the short propagation distance compared to the photon length, $z' \ll cT_r$, leads to the spectral representation $E_r(z',\tau) = e^{iz'\omega_r/c} \int_{-\infty}^{\infty} d\omega\, \tilde{E}'_r(\omega) e^{-i\omega\tau}$ in the form of the comb structure (Fig.3(a)),

$$\tilde{E}'_r(\omega) = \sum_{n=-\infty}^{\infty} \tilde{E}_n'(\omega) \text{ , where } \tilde{E}_n'(\omega) = \frac{1}{4\pi} \frac{E_0 e^{i\varphi_0} J_n(p) e^{in\vartheta_0}}{\Gamma_r + i(\omega_r - n\Omega - \omega)}. \tag{5}$$

Propagation of the electric field of the photon through the absorber, $E(z',\tau)$, $0 \leq z' \leq L$, with the boundary condition $E(z'=0,\tau) = E_r(z'=0,\tau)$ is described by the wave equation

$$\frac{\partial^2 E}{\partial z'^2} - \frac{1}{c'^2}\frac{\partial^2 E}{\partial \tau^2} = \frac{2\delta_e}{c'}\frac{\partial E}{\partial \tau} + \frac{4\pi}{\varepsilon c'^2}\frac{\partial^2 P}{\partial \tau^2}, \tag{6}$$

where $c' = c/\sqrt{\varepsilon}$, $\varepsilon \cong 1$ is the background dielectric permittivity of the absorber, $\delta_e$ takes into account the nonresonant losses due primaraly to photoelectric absorption, and $P'$ is the resonant nuclear polarization, expressed via the coherence of the absorber resonant nuclear transition as

$$P' = f_a N \cdot d_{12} \rho_{21}, \tag{7}$$

where $f_a$ is the probability of recoilless absorption, $N$ is the concentration of resonant nuclei, $\rho_{21}$ and $d_{12}$ is the coherence and the dipole moment of the resonant nuclear transition $|1\rangle$-$|2\rangle$, respectively (Fig.1(b)). Since the photons are depolarized and the dipole moments are randomly directed, one can omit the vector character of $E$, $P$, and $d_{12}$. The density matrix equation for the resonant coherence of the absorbing transition has the form

$$\frac{d\rho_{21}}{dt} + (i\omega_a + \gamma_a)\rho_{21} = \frac{i}{\hbar} n_{12} d_{21} \cdot E, \tag{8}$$

where $\omega_a$ is the frequency of the nuclear transition $|1\rangle$-$|2\rangle$, $\gamma_a$ is the halfwidth of the absorber spectral line, and $n_{12} = \rho_{11} - \rho_{22} \approx 1$ is the population difference between the energy levels $|1\rangle$ and $|2\rangle$.

We seek for a solution of (6)-(8) in the form $F(z',\tau) = e^{iz'\omega_r/c} \int_{-\infty}^{\infty} d\omega \tilde{F}'(z',\omega) e^{-i\omega\tau}$, where $F = \{E; P\}$. In the rotating wave approximation, $|\omega - \omega_a| \ll |\omega + \omega_a|$, we find from equations (7) and (8):

$$\tilde{P}'(z',\omega) = \frac{n_{12} f_a N |d_{12}|^2}{\hbar(\omega_a - \omega - i\gamma_a)} \tilde{E}'(z',\omega). \tag{9}$$

Substitution of (9) into (6) and implementation of the slowly-varying envelope approximation (implying the relations $|\partial \tilde{F}'(z',\omega)/\partial z'| \ll |\tilde{F}'(z',\omega)|\omega/c$ and $|\tilde{F}'(z',\omega)/\tilde{F}'(z',\omega_r)| \ll \omega_r/|\omega_r - \omega|$) results in the following equation:

$$\frac{\partial \tilde{E}'(z',\omega)}{\partial z'} + \left(\delta_e + \frac{T_M/L}{1+i(\omega_a-\omega)/\gamma_a}\right)\tilde{E}'(z',\omega) = 0, \tag{10}$$

where $T_M = f_a \dfrac{2\pi \omega N n_{12} |d_{12}|^2 L}{\gamma_a \hbar c \sqrt{\varepsilon}}$ is the Mössbauer optical thickness. Solution of (10), satisfying the input boundary condition $\tilde{E}'(z'=0,\omega) = \tilde{E}'_r(\omega)$, see (5), has the form of a comb of spectral components:

$$\tilde{E}'(z',\omega) = e^{-\delta_e z'} \sum_{n=-\infty}^{\infty} \tilde{E}_n'(\omega) e^{-\frac{T_M/L}{1+i(\omega_a-\omega)/\gamma_a}z'}. \tag{11}$$

at the output edge of the absorber This gives the time dependence of the output field behind the absorber both in the absorber reference frame, $E_{out}(z',\tau) = e^{iz'\omega_r/c} \int_{-\infty}^{\infty} d\omega \tilde{E}'(z'=L,\omega) e^{-i\omega\tau}$, and in the laboratory reference frame, $E_{out}(z,\tau)=e^{i\frac{\omega_r}{c}(z-R\sin(\Omega t))}\int_{-\infty}^{\infty}d\omega \tilde{E}'(z'=L,\omega)e^{-i\omega\tau}$. Finally, count rate of the transformed photons, $N_{out}(\tau) \sim |E_{out}(z,\tau)|^2$, has the form:

$$N_{out}(\tau) \sim e^{-2\delta_e L} E_0^2 \left| \int_{-\infty}^{\infty} d\omega e^{-i\omega\tau} \sum_{n=-\infty}^{\infty} \frac{J_n(p)e^{in\vartheta_0}}{\Gamma_r + i(\omega_r - n\Omega - \omega)} e^{-\frac{T_M}{1+i(\omega_a-\omega)/\gamma_a}} \right|^2. \quad (12)$$

Averaging of the output photon count rate, $N_{out}(\tau) = N_{out}(t-t_0)$, over the moment $t_0$ gives the dependence of the full flow of shaped gamma photons on the observation time.

$$N_{out}(t) = \lim_{T\to\infty}\frac{1}{T}\int_{-T}^{t}dt_0 N_{out}(t-t_0). \quad (13)$$

Substitution of (12) into (13) results in

$$N_{out}(t) \sim e^{-T_M} \sum_{n,m=-\infty}^{+\infty} J_n(p)J_m(p) B_{nm} \cos\{(n-m)(\Omega t + \vartheta_0) + \phi_{nm}\}, \quad (14)$$

where real parameters $B_{nm} = B_{mn}$ and $\phi_{nm} = \phi_{mn}$ are defined by the relation

$$B_{nm}e^{i\phi_{nm}} = \frac{1}{2\pi}\int_{-\infty}^{+\infty}\frac{1}{\omega^2+\Gamma_r^2} e^{-\frac{T_M}{2}\left(\frac{\gamma_a}{\gamma_a+i\Delta\omega_n}+\frac{\gamma_a}{\gamma_a-i\Delta\omega_m}\right)} d\omega, \quad (15)$$

$\Delta\omega_n = \omega_a - \omega_r + n\Omega - \omega$. The theoretical curves in Fig.3 and Fig.4 are plotted in accordance with (12) and (14)-(15), respectively. In order to fit the experimental data in Fig.3(b),(c), the time-dependencies were averaged over the acquisition intervals $\Omega dt$ were $dt$=5 ns determined by the experimental setup.

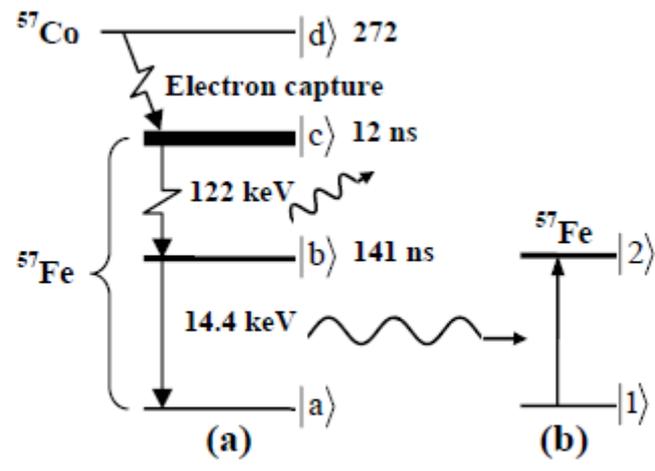

Fig. 1

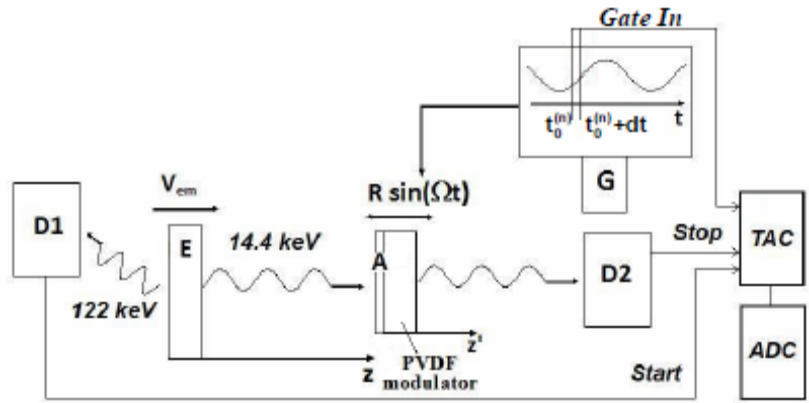

Fig. 2

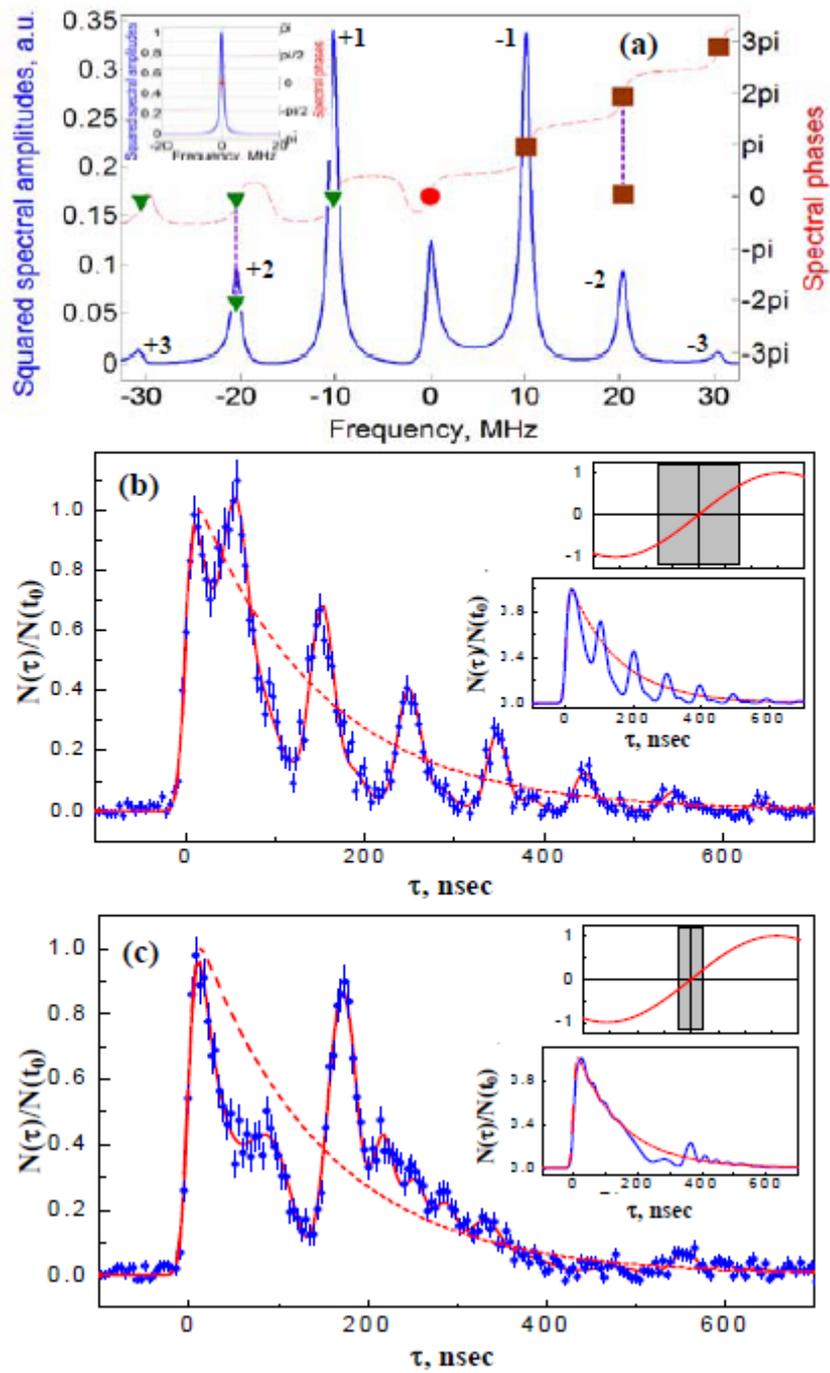

Fig. 3

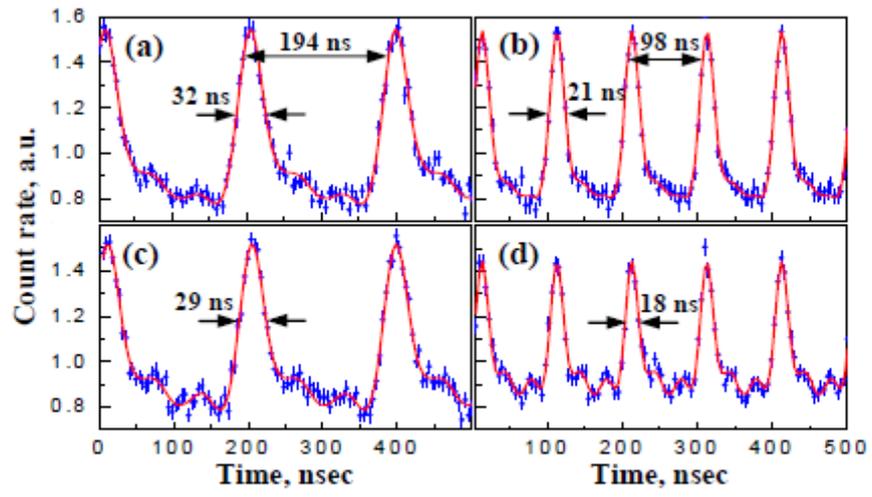

Fig. 4